\documentclass[conference]{IEEEtran}

\input epsf
\usepackage{graphicx}
\usepackage{cite}
\usepackage{amsmath,amssymb,amsfonts,amssymb}
\usepackage{graphicx}
\usepackage{textcomp}
\usepackage{xcolor}
\usepackage{mathrsfs}
\usepackage{caption}
\usepackage{array}
\allowdisplaybreaks  
\usepackage{graphicx}
\usepackage{verbatim}
\usepackage{enumitem}
\usepackage{booktabs}
\usepackage{tabularx}
\usepackage{bm}
\usepackage{color}
\usepackage{hyperref}
\hypersetup{hidelinks}

\makeatletter
\def\namedlabel#1#2{\begingroup
    #2%
    \def\@currentlabel{#2}%
    \phantomsection\label{#1}\endgroup
}


\newtheorem{remark}{Remark}
\newtheorem{definition}{Definition}
\newtheorem{theorem}{Theorem}

\newtheorem{example}{Example}

\newtheorem{lemma}{Lemma}

\usepackage{enumitem}
\usepackage[T1]{fontenc}
\usepackage[latin9]{inputenc}
\usepackage{amstext}
\usepackage{amssymb,amsmath}
\allowdisplaybreaks
\usepackage{graphicx}
\usepackage{caption}
\usepackage{subfigure}
\usepackage{verbatim} 
\usepackage[lined,ruled,commentsnumbered]{algorithm2e}
\usepackage{hyperref}

\newcommand{\F}{\mathbf{F}}
\newcommand{\x}{\mathbf{x}}
\newcommand{\y}{\mathbf{y}}
\newcommand{\I}{\mathbf{I}}

\newcommand{\p}{\mathbf{p}}
\newcommand{\q}{\mathbf{q}}

\newcommand{\R}{\mathbb{R}}

\newcommand{\sinr}{\textup{SINR}}

\begin{document}

\title{DIFFRACT: Neuralized Utility Maximization for Wireless Networks by Differentiable Programming}

\author{\IEEEauthorblockN{Chee Wei Tan\IEEEauthorrefmark{1},
Siya Chen\IEEEauthorrefmark{2}}
\IEEEauthorblockA{\IEEEauthorrefmark{1}\textit{Nanyang Technological University, Singapore, Nanyang Ave., Singapore} \\
\IEEEauthorrefmark{2}  \textit{Department of Computer Science, City University of Hong Kong, Hong Kong} \\
\IEEEauthorrefmark{1}\textit{cheewei.tan@ntu.edu.sg,} 
\IEEEauthorrefmark{2}  \textit{siyachen4-c@my.cityu.edu.hk} \\
}}

\maketitle

\begin{abstract}
Next-generation wireless networks, including satellite-to-Open RAN systems, demand agile and intelligent resource management capable of handling dynamic multi-user interference under stochastic quality of service constraints. This paper introduces DIFFRACT, a neuralized utility maximization framework that leverages differentiable programming to integrate deep learning with optimization in wireless networks. Central to our approach is the exploitation of the mathematical structure of standard interference functions, which are foundational in wireless power control. By developing a duality theory for these functions, we map iterative interference management algorithms into differentiable neural network architectures via algorithm unrolling. This enables distributed, end-to-end gradient-based learning at the network edge, supporting real-time adaptation to interference in both terrestrial and non-terrestrial environments. DIFFRACT allows for scalable and robust utility maximization by modeling complex channel dynamics and leveraging the expressiveness of differentiable models. Experimental results confirm the framework's theoretical soundness and practical effectiveness for next-generation wireless systems.
\end{abstract}
\IEEEoverridecommandlockouts

\IEEEpeerreviewmaketitle

\section{Introduction}

Lower satellite launch costs and growing global broadband demand have accelerated the commercialization of Low Earth orbit (LEO) satellites, spurring non-terrestrial network initiatives such as Starlink \cite{spacex1,spacex2}. These satellite-to-cell systems are poised to become integral components of next-generation wireless infrastructure, extending reliable connectivity to remote and underserved areas. At the same time, emerging wireless architectures such as 6G and Open Radio Access Networks (O-RAN) are being designed to support latency-sensitive and high-throughput applications by intelligently coordinating edge resources across heterogeneous terrestrial and non-terrestrial links \cite{spacex3,spacex4,spacex5}. The coexistence of diverse wireless interfaces and unpredictable interference poses fundamental challenges to distributed resource management. These complexities often lead to over-provisioned designs or underutilized spectrum, limiting network efficiency.

Deep learning has shown great potential for wireless resource optimization, offering the ability to learn adaptive policies directly from data in complex and dynamic environments such as non-terrestrial and satellite-to-cell networks, bypassing the need for explicit modeling of channels and interference \cite{osheadeeplearning, hoydis2024learning,hoydis2023sionna,sun2018learning, cui2020deep,chen2025openranet}.  Traditional optimization methods often struggle with intractable interference and nonconvex problem structures, whereas neural networks can generalize across a wide range of channel conditions and accelerate computation significantly through parallel processing on GPUs \cite{hoydis2023sionna}. Yet, deep learning alone often lacks the structure, interpretability, and convergence properties offered by classical theory.

This paper proposes DIFFRACT, a framework based on differentiable programming to neuralize utility maximization in uncertain, interference-limited wireless networks. By embedding the mathematical structure of standard interference functions into a differentiable computational graph, DIFFRACT enables gradient-based learning of power control policies that are both scalable and theoretically grounded. This approach is motivated by the emerging trend of {\it differentiable programming}, fueled by automatic differentiation in deep learning frameworks such as PyTorch and TensorFlow, which allows complex models to be trained end-to-end using gradient-based methods \cite{olahdp,edpbook, agrawal2020differentiating}. Inevitably, differentiable programming enables end-to-end optimization in AI-native wireless stacks that are GPU-compatible, e.g., NVIDIA's 6G Sionna \cite{hoydis2023sionna}.

In particular, we present new results rooted in the seminal standard interference function framework \cite{yates1995framework}, uncovering new links between monotone operator and an overlooked differentiability property, which we integrate into a broader differentiable programming paradigm. Our approach is based on advances in contractive interference mappings \cite{feyzmahdavian2012contractive}, log-convexity structures \cite{logconvex}, and differentiable convex programming frameworks such as LLCPs \cite{agrawal2020differentiating, agrawal2019differentiable}. These insights lead to DIFFRACT, a differentiable programming framework that embeds the structure of standard interference functions into computational graphs, enabling end-to-end learning through automatic differentiation tools like PyTorch and TensorFlow \cite{olahdp,edpbook, agrawal2020differentiating}. DIFFRACT is useful for satellite-to-cell networks with interference patterns due to atmospheric fading \cite{spacex9,spacex11,spacex12}, transforming interference-limited settings into learning systems with differentiable programming.

Moreover, the differentiability of these mappings offers programmable flexibility to accurately approximate complex real-world interference patterns, thereby enabling scalable, learning-based solutions for utility maximization in wireless networks. DIFFRACT leverages differentiable programming by integrating deep learning methods--such as algorithm unrolling \cite{gregor2010learning, monga2021algorithm,algorithmunrolling}--with fixed-point techniques. The duality framework not only guides the optimization but also offers a certificate of optimality, enabling distributed wireless network optimization through a ``{\it learn to optimize}" paradigm. Differentiable programming allows the embedding of iterative algorithms--such as those arising from traditional optimization problems -- directly into neural networks, where parameters can be trained through end-to-end gradient descent using automatic differentiation tools \cite{olahdp,edpbook, agrawal2020differentiating,algorithmunrolling}, which offers a data-driven alternative to classical optimization-theoretic methods. Furthermore, our study investigates how key design choices--such as the depth of neural networks and the degree of programmability in function approximation and unrolled algorithm iterations--affect the performance and generalization capabilities of DIFFRACT. This helps guide the development of scalable, GPU-compatible learning-based optimization strategies such as amortized optimization \cite{boydcvxmodel1, boydcvxmodel2}, advancing new differentiable neural solvers for wireless networks. The main contributions of this paper are:
\begin{enumerate}
\item We establish new results in duality and differentiability for the classical standard interference function, a foundational concept in wireless network resource allocation. These results enable differentiable programming for an AI-native optimization stack and guide the design of scalable, principled iterative solvers.

\item We develop a data-driven methodology to learn a broad class of implicit standard interference functions using neural networks. This approach leverages the derived primal-dual iterative algorithms and extends the applicability of standard interference models to a wider range of utility maximization tasks, enabling accurate modeling and fast adaptation in data-centric settings.

\item We introduce DIFFRACT, a deep learning architecture for scalable end-to-end differentiable programming. DIFFRACT efficiently learns interference models built on PyTorch and open-source differentiable proximal solver, supporting real-time deployment in wireless networks.
\end{enumerate}
  
This paper is organized as follows. In Section II, we introduce the wireless network model and the standard interference functions. In Section III, we develop the duality of standard interference functions that can be applied to utility maximization for wireless networks and optimally solved using primal and dual iterative algorithms. In Section VI, we propose a parameterized neural network-based framework for learning interference functions with illustrative examples. Then, we propose DIFFRACT to enable jointly distributed learning of interference functions and utility maximization that enhances the efficiency of training and distributed computation in Section V. Finally, we conclude the paper in Section VI. 

Notation: Bold uppercase letters denote matrices, bold lowercase letters denote vectors, italics denote scalars, and $\mathbf{u} \ge \mathbf{v}$ indicates componentwise vector inequality. Let $(\mathbf{B}\mathbf{u})_l$ denote the $l$th element of $\mathbf{B}\mathbf{u}$. Let $\mbox{diag}(\mathbf{u})$ be the diagonal matrix formed by $\mathbf{u}$. We write $\mathbf{B} \ge \mathbf{F}$ if $B_{ij} \ge F_{ij}$ for all $i,j$. The Perron-Frobenius eigenvalue of a nonnegative matrix $\mathbf{F}$ is denoted as $\rho(\mathbf{F})$, and the right and left eigenvector of $\mathbf{F}$ associated with $\rho(\mathbf{F})$ are denoted by $\mathbf{x}(\mathbf{F}) \ge \mathbf{0}$ and $\mathbf{y}(\mathbf{F}) \ge \mathbf{0}$ (or, simply $\mathbf{x}$ and $\mathbf{y}$, when the context is clear) respectively. Recall that the Perron-Frobenius eigenvalue of $\mathbf{F}$ is the eigenvalue with the largest absolute value. Assume that $\mathbf{F}$ is an irreducible nonnegative matrix.\footnote{A nonnegative matrix $\mathbf{F}$ is said to be irreducible if there exists a positive integer $m$ such that the matrix $\mathbf{F}^m$ has all entries positive.} Then $\rho(\mathbf{F})$ is simple and positive, and $\x(\mathbf{F}),\y(\mathbf{F}) > \mathbf{0}$ \cite{nonnegative_matrix}. The super-script $(\cdot)^{\top}$ denotes transpose. Denote $\mathbf{u}/\mathbf{v}$ by the vector $[u_1 / v_1, \dots, u_L / v_L]^{\top}$ and the Schur product between $\mathbf{x}$ and $\mathbf{y}$ by $\mathbf{x} \circ \mathbf{y} =\left(x_{1}y_{1}, \ldots, x_{L}y_{L}\right)^{\top}$. For any vector $\tilde\gamma=[\tilde\gamma_1,\ldots,\tilde\gamma_L]^{\top} \in\R^L$, let $e^{\tilde\gamma}=[e^{\tilde\gamma_1},\ldots,e^{\tilde\gamma_L}]^{\top}$, and $\log \mathbf{x}$ denotes $\log \mathbf{x}=\left(\log x_{1}, \ldots, \log x_{L}\right)^{\top}$.

\begin{figure*}
\centerline{\includegraphics[scale=0.38]{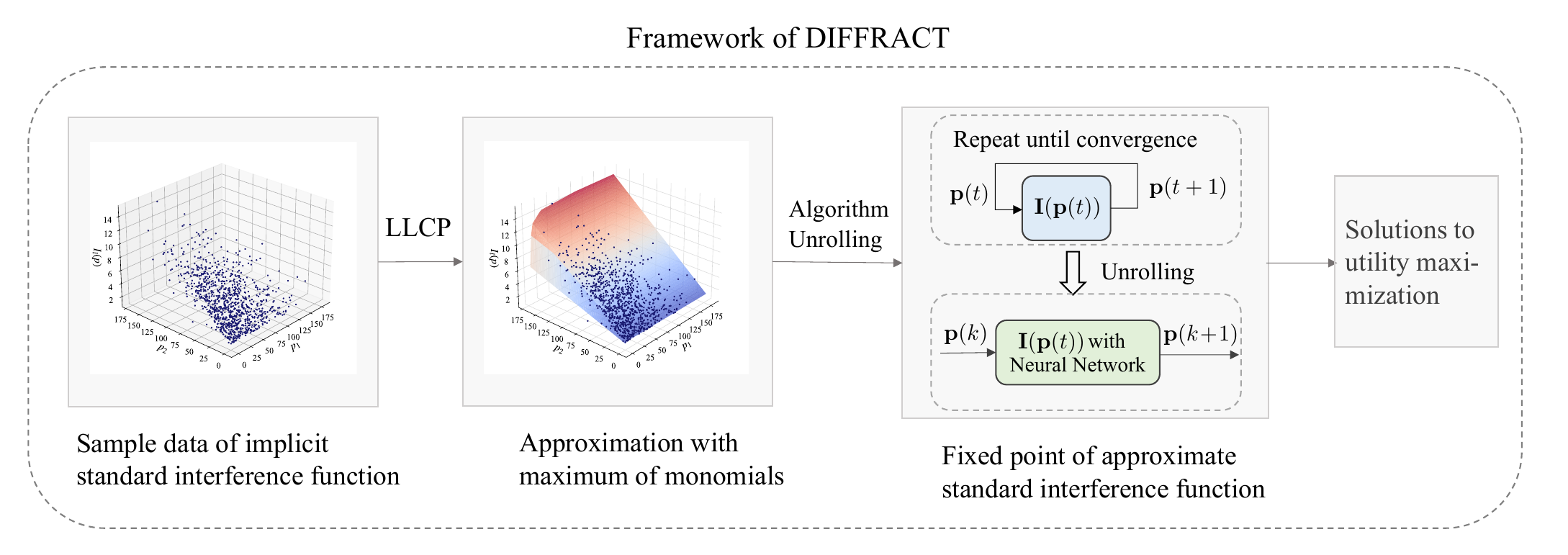}}
\captionsetup{font={footnotesize}, justification=raggedright}
\caption{DIFFRACT as a differentiable programming framework for utility optimization approximates implicit interference functions via differentiable log-convex mapping and uses algorithm unrolling to compute fixed points, enabling end-to-end learning of globally optimal solutions.\label{fig:unrolling}}
\end{figure*}

\section{Interference Framework and System Model}

\subsection{Standard Interference Function}
We revisit the standard interference function framework from \cite{yates1995framework}, revealing its underlying monotone operator structure and differentiability properties. This enables a new class of data-driven algorithms via differentiable programming.
\begin{definition}[Standard Interference Function \cite{yates1995framework}]\label{def:sif}$\mathbf{I}(\mathbf{p})$ is a standard interference function if, for all $\mathbf{p}\geq 0$, the following properties are satisfied:
\begin{enumerate}
    \item (Monotonicity) If $\mathbf{p}_1\geq \mathbf{p}_2$, then $\mathbf{I}(\mathbf{p}_1)\geq \mathbf{I}(\mathbf{p}_2)$.
    \item (Scalability) For any $\alpha>1, \alpha\mathbf{I}(\mathbf{p})>\mathbf{I}(\alpha\mathbf{p})$.
\end{enumerate}
\end{definition}
\begin{lemma}
\label{lemma2}\cite{yates1995framework}
If $\I(\p)\leq \p$ is feasible, $\I(\p)$ has a unique fixed point $\p^*$, which can be computed by the iteration $\p(t+1) = \I(\p(t))$ with geometric convergence rate.
\end{lemma}
\begin{lemma}\label{lemma1}
Positivity and concavity of $\mathbf{I}(\p)$ implies scalability of $\mathbf{I}(\p)$. 
\end{lemma}

To see this, note that for all $\alpha>1$, we have $\alpha\mathbf{I}(\bm{p})= \alpha\mathbf{I}(1/\alpha\cdot (\alpha\bm{p})+(1-1/\alpha)\cdot\bm{0})
    \geq \alpha\left((1/\alpha)\cdot\mathbf{I}(\alpha\bm{p})+(1-1/\alpha)\cdot\mathbf{I}(\bm{0})\right)
    > \alpha\cdot(1/\alpha)\cdot\mathbf{I}(\alpha\bm{p})+\bm{0}
    = \mathbf{I}(\alpha\bm{p})$ where the last two inequalities are due to concavity and positivity respectively.
However, the converse does not hold; that is, positivity and scalability do not imply concavity. A counterexample is $\mathbf{I}_i(\mathbf{p})=p_i^2 + 3$ if $0\leq p_i \leq 1$ and $\mathbf{I}_i(\mathbf{p}) =2p_i + 2$ if $p_i > 1$.

We first state a new result for the standard interference function framework in \cite{yates1995framework} with applications to utility maximization in the next section. Suppose that $\I(\p)$ is differentiable and concave in $\p$. Denote the derivative of $\I(\p)$ by:
\begin{equation}
    \nabla \I(\p) = \left(\frac{\partial I_i(\p)}{\partial p_j}\right)_{L\times L}.
\end{equation}
Observe that $\nabla \I(\p)$ is an irreducible nonnegative matrix whose entries are nonnegative and continuous in $\mathbf{p}$. The feasibility of $\I(\p) \le \p$ can be established by the following necessary and sufficient condition.
\begin{theorem}\label{lemma3}
If $\I(\p)$ is differentiable and concave, then $\I(\p)\leq \p$ is feasible if and only if $\rho(\nabla \I(\p))< 1$ for some $\p \geq \bm{0}$.
\end{theorem}  

\begin{definition}\label{def:logconvex}
$\I(\mathbf{p})$ is a \textit{log-log convex standard interference function}, if it fulfills the condition of standard interference function and, in addition, in the logarithmic domain, i.e., $\tilde{\mathbf{p}}=\log \mathbf{p}$, $\I(e^{\tilde{\mathbf{p}}})$ is log-convex in $\tilde{\p}$.
\end{definition}

We study a broad class of concave and log-log convex standard interference functions that subsume all known cases and underpin wireless utility maximization  \cite{yates1995framework,Chiang08}. Notably, as $\I(e^{\tilde{\mathbf{p}}})$ is convex and monotone, $\I(e^{\tilde{\mathbf{p}}})$ can be treated as a {\it monotone operator} established as follows.
\begin{theorem}[Interference Function as Monotone Operator]
\label{lemma4}
Let $I: \mathbb{R}^L_{++} \to \mathbb{R}^L_{++}$ be a standard interference function that is continuously differentiable, positive, monotone and scalable. Define $\Phi: \mathbb{R}^L \to \mathbb{R}^L$ by
$\Phi(\tilde{\mathbf{p}}) = \tilde{\mathbf{p}} - \log \I(e^{\tilde{\mathbf{p}}})$. Then $\Phi$ is monotone (e.g., see \cite{bauschke2017convex}), i.e., $\left( \Phi(\tilde{\mathbf{p}}_1) - \Phi(\tilde{\mathbf{p}}_2) \right)^{\top} \left( \tilde{\mathbf{p}}_1 - \tilde{\mathbf{p}}_2 \right) \geq 0$ for all $\tilde{\mathbf{p}}_1, \tilde{\mathbf{p}}_2 \in \mathbb{R}^L$.
\end{theorem}

Let us sketch the proof outline. Define $\Psi(\tilde{\mathbf{p}}) = \log \I(e^{\tilde{\mathbf{p}}})$, so $\Phi(\tilde{\mathbf{p}}) = \tilde{\mathbf{p}} - \Psi(\tilde{\mathbf{p}})$. The Jacobian is
\[
\nabla \Phi(\tilde{\mathbf{p}}) = I_L - \nabla \Psi(\tilde{\mathbf{p}}),
\]
where $I_L$ is identity matrix, and by the chain rule with $\mathbf{q} = e^{\tilde{\mathbf{p}}}$,
\[
\nabla \Psi(\tilde{\mathbf{p}}) = \text{diag}(\I(\mathbf{q}))^{-1} \cdot \nabla \I(\mathbf{q}) \cdot \text{diag}(\mathbf{q}).
\]
Now, the symmetric part of $\nabla \Phi$ is $\frac{1}{2} (\nabla \Phi + \nabla \Phi^T) = I_L - \frac{1}{2} (\nabla \Psi + \nabla \Psi^T)$. Since $\I(\mathbf{q})$ is monotone and differentiable, $\nabla \I(\mathbf{q})$ is positive semi-definite, and $\text{diag}(\I(\mathbf{q}))^{-1}$, $\text{diag}(\mathbf{q})$ are positive definite. Thus, $\nabla \Psi$ has a positive semi-definite symmetric part, so $I_L - \frac{1}{2} (\nabla \Psi + \nabla \Psi^T)$ is positive semi-definite, implying that $\Phi$ is monotone, i.e., $\left( \Phi(\tilde{\mathbf{p}}_1) - \Phi(\tilde{\mathbf{p}}_2) \right)^{\top} \left( \tilde{\mathbf{p}}_1 - \tilde{\mathbf{p}}_2 \right) \geq 0$ for all $\tilde{\mathbf{p}}_1, \tilde{\mathbf{p}}_2 \in \mathbb{R}^L$ \cite{bauschke2017convex}. 

Theorem \ref{lemma4} enables the use of operator splitting techniques such as proximal point or forward-backward methods in \cite{beck2017firstorder,bauschke2017convex} to solve equations of the form $\tilde{\mathbf{p}}^{\star} = \log \I(e^{\tilde{\mathbf{p}}^{\star}})$ with guaranteed convergence. An example is Nesterov's adaptive step-size rule update \cite{beck2017firstorder}: 
\begin{equation}
\label{nesterovfirstorder}
    \tilde{\mathbf{p}}(k+1) = \tilde{\mathbf{p}}(k) + \alpha_k(\log \I(e^{\tilde{\mathbf{p}}(k)}) - \tilde{\mathbf{p}}(k)),
\end{equation}
with $\alpha_k = \frac{2}{1 + \rho(\nabla \I(e^{\tilde{\mathbf{p}}(k)})}$ where the distance between one and the Perron-Frobenius eigenvalue of the Jacobian, $\delta$ (i.e., $1 - \rho(\nabla \I(e^{\tilde{\mathbf{p}}}))$ as the Jacobian is nonnegative) governs the convergence rate \cite{beck2017firstorder,bauschke2017convex}. The error after $k$ iterations of (\ref{nesterovfirstorder}) satisfies the bound \cite{beck2017firstorder,bauschke2017convex}:$
\| \tilde{\mathbf{p}}(k) - \tilde{\mathbf{p}}^{\star} \|_2 \leq C (1 - \delta)^k = C \rho(\nabla \I(e^{\tilde{\mathbf{p}}(k)}))^k,
$ for some positive constant $C$ determined by the initialization. This is as competitive as the fixed-point iteration in Lemma \ref{lemma2}! This operator perspective enables differentiable programming \cite{edpbook} to address a broad class of wireless network optimization efficiently, as shown in Fig.~\ref{fig:unrolling}.

\subsection{System Model}
Consider the system model of a wireless network with $L$ users, each comprising a transmitter-receiver pair, communicating simultaneously over a shared spectrum. The vector $\p = (p_1, \dots, p_L)^\top$ denotes the allocation of transmit power, where $p_l$ represents the transmit power of the $l$th user. Let $\mathbf{G}=[G_{lj}]_{l,j=1}^L>0_{L\times L}$ represent the wireless channel gain, where $G_{lj}$ is the channel gain from the $j$th transmitter to the $l$th receiver, and $n=(n_1, \dots, n_L)^\top >\mathbf{0}$, where $n_l$ is the noise power at the $l$th user. We define the Signal-to-Interference-and-Noise Ratio (SINR) at the $l$th receiver as the ratio of the received signal power to the combined interference and noise power under frequency-flat fading, given by:
\begin{equation}\label{eq:sinrp}
\textup{SINR}_l(\p) = \displaystyle \frac{G_{ll}p_l }{\sum_{j=1,j \neq l}^L G_{lj}p_j + n_l}.
\end{equation}

Next, let us define a  nonnegative vector:
$$
\mathbf{v}= \left(\frac{n_1}{G_{11}}, \frac{n_2}{G_{22}}, \dots,\frac{n_L}{G_{LL}}\right)^{T},
$$
and a nonnegative matrix $\F$ with entries:
\begin{equation}\label{eq:matrixF}
F_{lj}=\left\{\begin{matrix}
0, &\mbox{if}\;\; l = j \\ 
G_{lj}/G_{ll}, &\mbox{if}\;\; l \ne j
\end{matrix}\right..
\end{equation}
Moreover, we assume that $\F$ is irreducible, i.e., each link has at least one interferer. Then the SINR of the $l$th user can be rewritten as:
\begin{equation}\label{eq:sinrp}
\textup{SINR}_l(\p) = \displaystyle \frac{ p_l }{\sum_{j=1}^L F_{lj}p_j +v_l}.
\end{equation}
Suppose each $\textup{SINR}l(\p)$ must exceed a reliability threshold $\gamma_l$, i.e., $\textup{SINR}l(\p)\geq \gamma_l$. This constraint induces the well-known standard interference function \cite{yates1995framework}:
\begin{align}\label{eq:if_sinr}
\I_l(\p)=\sum_{j=1}^L \gamma_lF_{lj}p_j +\gamma_lv_l \leq p_l \quad \forall, l,
\end{align}
which is feasible if and only if $\rho(\text{diag}(\boldsymbol{\gamma})\mathbf{F})<1$, as established in Theorem \ref{lemma3}. When the wireless channels experience fading (e.g., Rayleigh fading, Rician fading or Nakagami fading), the received power from the $j$th transmitter at the $l$th receiver is given by $G_{l j} h_{l j} p_{j}$ where $h_{l j}$ is a random variable reflecting the fading environment. The SINR of the $l$th user is a random variable in terms of the channel realization:
\begin{equation}\label{eq:sinr2}
\textup{SINR}_l(\p) = \displaystyle \frac{h_{ll}p_l }{\sum_{j=1}^L F_{lj}h_{lj}p_j +v_l}.
\end{equation}
An outage occurs when the received SINR of the $l$th user falls below $\gamma_l$, a minimum SINR threshold for reliable communication. This means that when $\operatorname{SINR}_{l}(\mathbf{p}) \geq \gamma_l$, the transmission at the $l$th receiver is successful; otherwise, the transmission fails. Denote the outage probability of the $l$th receiver/transmitter pair for a power vector $\p$ by $ O_l(\p)$ and the outage constraint for the $l$th user by $\bar{O}_l$. The link reliability function of the $l$th user is given as the complement of the outage probability:
\begin{align}\label{eq:con_out}
     O_l(\p) =\operatorname{Prob}(\sinr_l(\p)<\gamma_l)\leq \bar{O}_l.
\end{align}

While a closed-form standard interference function expression for \eqref{eq:con_out} is available under Rayleigh fading (e.g., see \cite{tan2015optimal}), such expressions are generally intractable for other fading conditions, particularly atmospheric fading in space-to-ground communication scenarios \cite{spacex9,spacex11,spacex12}. This motivates the development of a framework that can algorithmically handle implicit interference functions for utility maximization under these more complex channel conditions.

\section{Utility Maximization with log-log convex Interference Function}\label{sec:sec3}
In this section, we study a utility maximization problem with log-log convex differentiable interference function constraints. We adopt an operator perspective by exploiting the problem's duality, which yields fast fixed-point algorithms. Differentiability of these operators then enables a differentiable programming framework to learning implicit models and paving the way for {\it neuralized utility maximization}.

\subsection{Utility Maximization and Iterative Algorithm}
Let $u(\p)$ denote the network utility, representing a metric of network quality-of-service. The power vector $\p = [p_1, \ldots, p_L]^{\top}$ is adjusted to maximize this utility, subject to interference constraints $\I(\p) \leq \p$. Consider the wireless utility maximization problem:
\begin{align}\label{eq:prob_cvx}
\mbox{maximize}\;\;\;\; & u(\p) \nonumber\\
\mbox{subject to}\;\;\;\; & \I(\p) \leq \p.
\end{align}

Letting $\tilde{\p} = \log \p$, we assume that the utility function $u(e^{\tilde{\p}})$ is concave and monotonically increasing in terms of $\tilde{\p}$. One example is the minimum of a series of competitive utility functions that satisfy the following assumptions \cite{zheng2016wireless}. 
\begin{definition}[Competitive Utility Functions \cite{zheng2016wireless}]\label{assm1}
\begin{enumerate}
    \item {\it Competitiveness:} For all $i$, the utility $u_i(\p)$ is strictly increasing with respect to $p_i$ and strictly decreasing with respect to $p_j$ for all $j \neq i$, whenever $p_i > 0$.
    \item {\it Directional Monotonicity:} For any $s > 1$ and $\mathbf{p} > \mathbf{0}$, $u_i(s \mathbf{p}) > u_i(\mathbf{p})$ for all $i$.
    \item $u_i(e^{\tilde{\p}})$ is concave and monotonically increasing for all $i$.
\end{enumerate}
\end{definition}



Now, taking a logarithmic transformation of the variables, \eqref{eq:prob_cvx} can be reformulated as: 
\begin{align}
\label{eq:prob_ecvx}
\mbox{maximize}\;\;\;\; & u(e^{\tilde{\p}}) \nonumber\\
\mbox{subject to}\;\;\;\; & \log(\I(e^{\tilde{\p}})) \leq \tilde{\p},
\end{align}
which reduces to a convex optimization problem, thus allowing (\ref{eq:prob_cvx}) to be solved optimally. Suppose the constraints in \eqref{eq:prob_cvx} are feasible and there exists an $\tilde{\p}$ that is strictly feasible, i.e., Slater's condition ~\cite{Convex_Optimization} is satisfied. Let us introduce the dual variable $\bm{\lambda} \in \mathbb{R}_{+}^{L}$ and form the partial Lagrangian of \eqref{eq:prob_ecvx} as:
$$\mathcal{L}(\tilde{\p}, \bm{\lambda}) = -u(e^{\tilde{\p}}) + \bm{\lambda}^{\top}(\log(\I(e^{\tilde{\p}})) - \tilde{\p}).$$
Applying the  Karush-Kuhn-Tucker (KKT) stationarity condition~\cite{Convex_Optimization} and transforming back to the original variables, we obtain the following key result.
\begin{theorem}\label{thm:thm1}
The optimal solution $\p^*$ of \eqref{eq:prob_cvx} and the optimal dual solution $\bm{\lambda}^*$ of \eqref{eq:prob_ecvx} satisfy:
\begin{align}\label{eq:kkt}
    \frac{\lambda_l^*}{p_l^*} = -\frac{\partial u(\p^*)}{\partial p_l} + \sum_{j=1}^{L}\frac{\lambda_j^*}{p_j^*}\frac{\partial I_j(\p^*)}{\partial p_l} \;\; \forall \; l.
\end{align}
Furthermore, the following iterations compute $\p^*$ in (\ref{eq:prob_cvx}) and $\bm{\lambda}^*$ in (\ref{eq:prob_ecvx}) from any positive initial points $\p(0)$ and $\x(0)$:
\begin{align}\label{eq:eq1}
    \p(t+1) = \I(\p(t)),
\end{align}
\begin{align}\label{eq:eq2}
\x(t+1)= - \nabla u(\p(t)) + \nabla \I(\p(t))\x(t),
\end{align}
and
$$\bm{\lambda}(t+1) = \p(t+1)\circ \x(t+1)$$
if and only if $\rho(\nabla \I(\p))< 1$ for some $\p\geq \bm{0}$.
\end{theorem}

\begin{remark}
The fact that the spectral radius of the Jacobian matrix $\nabla \I(\p(t))$ is strictly less than one for any $\p(t)$ (as shown in Theorem \ref{lemma3}), has been established in \cite{abara2018spectral} from a control-theoretic perspective based on studying the existence of a fixed point and its convergence behaviour, which are established using the contractive interference function findings in \cite{feyzmahdavian2012contractive,feyzmahdavian2014stability}. Theorem \ref{thm:thm1} offers an alternative and practical interpretation of this result by linking Theorem \ref{lemma3} to the use of iterative algorithms for solving optimization problems.
\end{remark}

\subsection{Convex Relaxation to Utility Maximization}
When the problem~\eqref{eq:prob_cvx} is infeasible, we consider an approximate power solution that relaxes the original constraints. Specifically, we introduce the following relaxation of~\eqref{eq:prob_cvx}:
\begin{align}
\label{eq:prob_relax}
\mbox{maximize}\;\;\;\; & u(\p) \nonumber\\
\mbox{subject to}\;\;\;\; & \prod_{l=1}^L\left(\frac{I_l(\p)}{{p}_l}\right)^{w_l}\leq 1,
\end{align}
where $\bm{w} = [w_1,\cdots,w_L]^{\top}$ is a positive vector that is a convex combination of the individual constraints in \eqref{eq:prob_cvx} and satisfying $\sum_{l=1}^L w_l = 1$ and $w_l\geq 0$. By letting $\tilde{\p} = \log \p$, \eqref{eq:prob_relax} can be solved by the following convex optimization problem:
\begin{align}\label{eq:prob_relax2}
\mbox{maximize}\;\;\;\; & u(e^{\tilde{\p}}) \nonumber\\
\mbox{subject to}\;\;\;\; & \sum_{l=1}^L w_l \left( \log I_l(e^{\tilde{\p}})- \tilde{p}_l \right) \leq 0.
\end{align}


\begin{remark}
The term $\prod_{l=1}^L \bigl(I_l(\p)\bigr)^{w_l}$ in~\eqref{eq:prob_relax} is concave as it is the weighted geometric mean of concave nonnegative functions, which preserves concavity under vector composition~\cite{Convex_Optimization}. If~\eqref{eq:prob_cvx} is indeed feasible, then by Theorem~\ref{lemma3} and Theorem~\ref{thm:thm1}, setting $\mathbf{w}=\mathbf{p}^{\star}\circ\mathbf{x}^{\star}$ in~\eqref{eq:prob_relax} recovers the solution $\mathbf{p}^{\star}$ of~\eqref{eq:prob_cvx}, and the relaxation is tight.
\end{remark}


\subsection{Relationship to Nonlinear Perron-Frobenius Theory}
We remark that an interesting duality arises between the algorithms for computing 
$\mathbf{p}$ and $\mathbf{x}$, which can be understood via nonlinear Perron--Frobenius theory \cite{tan2015wireless,zheng2016wireless,zheng2017max,keener1993perron,krause2001concave}: 
the vectors $\mathbf{p}$ and $\mathbf{x}$ can be viewed as Perron--Frobenius eigenvectors of suitably constructed nonlinear operators.
\begin{theorem}[Perron--Frobenius Theorem Characterization]
\label{npftheorem}
Let $\mathbf{I}:\mathbb{R}^L_{+}\!\to\!\mathbb{R}^L_{+}$ be a differentiable concave standard interference mapping and let $u:\mathbb{R}^L_{+}\!\to\!\mathbb{R}$ be a differentiable utility function. Suppose $\mathbf{p}^\star$ satisfies $\mathbf{p}^\star = \mathbf{I}(\mathbf{p}^\star)$ and $\rho(\nabla \mathbf{I}(\mathbf{p}^\star)) < 1$. By nonlinear Perron--Frobenius theory, such a map is nonexpansive in Hilbert's projective metric
$d_H(\p,\q) := \log\frac{\max_i p_i/q_i}{\min_i p_i/q_i}$, and admits a unique fixed point $p^\star$ up to scaling, which minimizes the projective diameter of the orbit of $\mathbf{I}(\cdot)$. The fixed point $\p^\star = \mathbf{I}(\p^\star)$ plays the role of the nonlinear Perron--Frobenius eigenvector, characterized by a Collatz--Wielandt--type formula
$$
\inf_{\mathbf{p} > \mathbf{0}} \max_i \frac{I_i(\mathbf{p})}{p_i} \;=\; \sup_{\mathbf{p} > 0} \min_i \frac{I_i(\mathbf{p})}{p_i} \;=\; 1.
$$
Furthermore, $\mathbf{x}^\star$ in Theorem \ref{thm:thm1} satisfies
$\mathbf{x}^\star = \nabla \mathbf{I}(\mathbf{p}^\star)\, \mathbf{x}^\star - \nabla u(\mathbf{p}^\star)$, or equivalently,
\[
\mathbf{x}^\star = \bigl(\bm{I}_L - \nabla \mathbf{I}(\mathbf{p}^\star)\bigr)^{-1}\bigl(-\nabla u(\mathbf{p}^\star)\bigr).
\]
Moreover, since $\mathbf{I}(\cdot)$ is concave, the Jacobian mapping $\mathbf{p} \mapsto \nabla \mathbf{I}(\mathbf{p})$ is order-preserving and concave, and the induced operator $\mathbf{x} \mapsto \nabla \mathbf{I}(\mathbf{p}^\star) \mathbf{x}$ belongs to the class of concave Perron--Frobenius operators. Consequently, $\mathbf{x}^\star$ lies in the Perron--Frobenius cone of $\nabla \mathbf{I}(\mathbf{p}^\star)$.
\end{theorem}

The special case where $\nabla u(\mathbf{p}^\star)= - \mathbf{w}^T\mathbf{p}$ and $\mathbf{I}(\mathbf{p})=\mathbf{F}\mathbf{p}+ \mathbf{v}$ leads to a solution $\mathbf{p}^\star$ and $\mathbf{x}^\star$ that recovers the {\it uplink-downlink duality} and yields a PageRank-like structure, as shown in \cite{TON09}. Also, the Collatz-Wielandt-type objective function is a restatement of scalability in Definition 1 \cite{riedel2025deep}.

\begin{remark}
    Since $\I(\bm{e}^{\tilde{\p}})$ is convex in $\tilde{\mathbf{p}}$, its Jacobian matrix 
$J(\tilde{\mathbf{p}}) := \frac{\partial I_i(\tilde{\mathbf{p}})}{\partial \tilde{p}_j}$
has nonnegative entries and satisfies
$$
J(\tilde{\mathbf{p}}) = \operatorname{diag}\big(\I(e^{\tilde{\mathbf{p}}})\big)^{-1} \nabla \I(e^{\tilde{\mathbf{p}}}) \operatorname{diag}\big(e^{\tilde{\mathbf{p}}}\big),
$$
where $\nabla \I(\cdot)$, characterized in Theorems \ref{lemma3} and \ref{lemma4}, enables efficient automatic differentiation techniques studied next.
\end{remark}

\section{DIFFRACT: LLCP Modeling and Algorithm Unrolling for Utility Maximization}\label{sec:sec4}

\begin{figure}
\centerline{\includegraphics[scale=0.35]{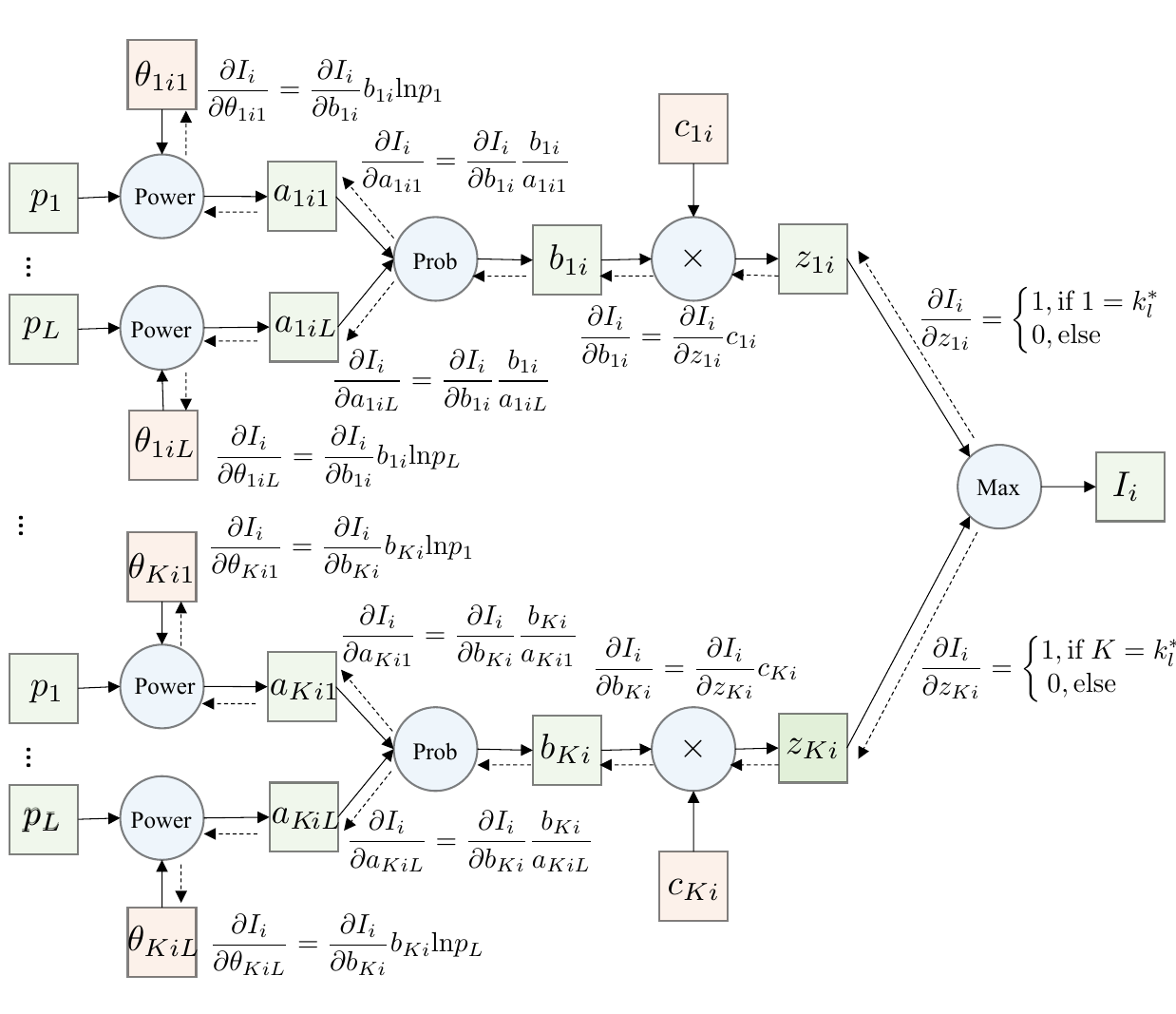}}
\captionsetup{font={footnotesize}, justification=raggedright}
\caption{Computational graph of the DIFFRACT framework, implementable in PyTorch with automatic differentiation, for modeling standard interference functions (potentially non-smooth) by using Lemma \ref{lemma5} and Theorem \ref{lemma2}.\label{fig:compugraph}}
\end{figure}

While the standard interference function framework enables fast-converging iterative algorithms, this is limited to deterministic closed-form expressions. To overcome this limitation, we must extend the framework to accommodate {\it implicit standard interference functions}---those that cannot be expressed in deterministic form, such as in the case of general fading in \eqref{eq:con_out}. To address this, we propose a data-driven framework called DIFFRACT. Specifically, we approximate a log-log convex but implicit standard interference function using the LLCP model in \cite{agrawal2020differentiating}, converting it into explicit expressions with monomials. We then employ algorithm unrolling \cite{gregor2010learning,monga2021algorithm} to transform the fixed-point iterations into a fixed-depth neural network, whose parameters are then data-optimized to maximize computational efficiency. The architecture of DIFFRACT is shown in Fig.~\ref{fig:unrolling}.


\subsection{Modeling Standard Interference Functions with LLCP for Stochastic Outage Probability}
To approximate the implicit standard interference function in \eqref{eq:con_out}, we use a data-driven method in \cite{agrawal2020differentiating} to learn a log-log convex function from data, leveraging a monomial basis for its approximation as given by the following lemma. The approximation accuracy is then demonstrated through illustrative examples with validation against ground truth.
\begin{lemma}\label{lemma5}
Consider the standard interference function 
\begin{align}\label{eq:mv}
    &\I(\p)\!\\\nonumber
    =&\!\!\left[\max_{k=1,\cdots,K} \left\{c_{k1}\prod_{j=1}^{L}p_j^{\theta_{k1j}}\right\}, \cdots, \!\max_{k=1,\cdots,K}\left\{c_{kL}\prod_{j=1}^{L}p_j^{\theta_{kLj}}\right\}\!\right]
\end{align}
where $c_{kl}\geq 0,\theta_{klj}\geq 0$. Suppose $\sum_{j=1}^L\theta_{klj}+\epsilon\leq
1$ for all $l,k$ for some positive $\epsilon$, $\I(\p)$ is feasible and log-log convex.
\end{lemma}

We learn the parameters $c_{kl}$ and $\theta_{klj}$ in \eqref{eq:mv} from training data via the LLCP model in \cite{agrawal2020differentiating}:
\begin{align}\label{opt:cvxL}
\hat{\y} (\p)=\textup{argmin}_{y}\;\;\;\;&\textbf{1}^{\top}(\bm{z}/\y+\y/\bm{z})\nonumber\\
\mathrm{subject\;to}&\;\; z_l = \!\max_{k=1,\cdots,K} \left\{c_{kl}\prod_{j=1}^{L}p_j^{\theta_{klj}}\right\}, \forall \, l\nonumber\\
\mathrm{variables:} & \;\;\;\y,\bm{z}\in \mathbb{R}^L_+,
\end{align}
where $c_{kl}$ and $\theta_{klj}$ are the learning parameters, and $p_j$ denotes the input of the training data to \eqref{opt:cvxL}. Suppose we have a training dataset $\bm{\mathcal{D}}$ consisting of $ |\bm{\mathcal{D}}|$ input-output pairs $\{ \p(i), \y(i) \} \in \bm{\mathcal{D}}$. We evaluate the performance of \eqref{opt:cvxL} on this dataset using the following mean squared error loss:
\begin{align}\label{eq:loss}
E(\bm{c},\bm{\theta})=&M\sum_{k=1}^K\sum_{l=1}^L \phi(\sum_{j=1}^{L} \theta_{klj}-1)\nonumber\\
    & \quad +\frac{1}{|D|}\sum_{\{\p(i),\y(i)\}\in \bm{\mathcal{D}}}\|\y-\hat{\y} (\p(i))\|_2^2,
\end{align}
where $\bm{c}= (c_{kl})_{K\times L}$, $ \bm{\theta}= (\theta_{klj})_{K\times L\times L}$, $M$ is a large positive number and $\phi(\cdot)$ is a penalty function, e.g., $\phi(\sum_{j=1}^{L}\theta_{klj}-1) = \left( \max\left\{0,\sum_{j=1}^{L}\theta_{klj}-1\right\} \right)^2$. The first term of \eqref{eq:loss} is differentiable (due to the square and despite $\max$) and ensures that $\sum_{j=1}^L\theta_{klj}<1$, i.e., $\I(\p) \le \p$ is feasible. We minimize the training loss $E(\bm{c},\bm{\theta})$ via gradient descent, with gradients obtained through automatic differentiation (e.g., {\tt PyTorch autograd}); Fig.~\ref{fig:compugraph} illustrates the resulting computational graph. This is summarized in Algorithm \ref{alg:alg1}.

\begin{algorithm}
\SetAlgoLined
\textbf{Input}: Training dataset $\bm{\mathcal{D}}$ with input-output pairs $(\p{(i)},\y{(i)})\in \bm{\mathcal{D}}.$\\
\textbf{Initialize}: Learning parameters $\bm{c}{(0)}$ and $\bm{\theta}{(0)}$, learning rate $\eta_c$ and $\eta_{\theta}$.\\
\For{$i = 1,\dots,|\bm{\mathcal{D}}|$}{
Obtain $\hat{\y} (\p{(i)})$ by solving \eqref{opt:cvxL} with $\bm{c}{(i)}$, $\bm{\theta}{(i)}$ and $\p{(i)}$.\\
Use automatic differentiation (e.g., {\tt Pytorch autograd}) to compute the gradients $\nabla_c E(\bm{c}{(i)},\bm{\theta}{(i)})$ and $\nabla_{\theta} E(\bm{c}(i),\bm{\theta}(i))$.\\
Update the parameters $\bm{c}{(i+1)} = \bm{c}{(i)}-\eta_c\nabla_c E(\bm{c}{(i)},\bm{\theta}{(i)})$ and  $\bm{\theta}{(i+1)} = \bm{\theta}{(i)}-\eta_{\theta}\nabla_{\theta}E(\bm{c}{(i)},\bm{\theta}{(i)})$.
}
\textbf{Output}: Monomial approximation $\I(\p)$ in \eqref{eq:mv} with learned $ \bm{c} $ and $ \bm{\theta}$.
\caption{Monomial approximation via automatic differentiation.}
\label{alg:alg1}
\end{algorithm}


We evaluate the effectiveness of the LLCP model in \eqref{opt:cvxL} in approximating the implicit function $\I(\p)$ in \eqref{eq:con_out} with explicit expressions under Rayleigh and Rician fading models relevant to satellite-terrestrial networks \cite{stuber1996principles, spacex9, spacex10, spacex11}. Interestingly, Rayleigh fading allows closed-form deterministic functions, hence enabling direct validation of the LLCP model in \eqref{opt:cvxL}. In contrast, Ricean fading lacks exact expressions, but can still be effectively approximated using the same model.

\subsubsection{Modeling Interference Function for Outage Constraints under Rayleigh fading}

Suppose both the desired signals and
interference signals are subject to Rayleigh fading, then $h_{l j}$ are independent and exponentially distributed with unit mean, i.e., $E\left[F_{l j} h_{l j} p_{j}\right]=F_{l j} p_{j}$ \cite{boyd02}. The outage probability of the $i$th user can be written in closed-form as \cite{boyd02}:
\begin{align}\label{eq:outage1}
O_l(\p) &=\textup{Prob}(\sinr_l(\p)<\gamma_l)\nonumber\\
&= 1-e^{\frac{-\gamma_lv_l}{p_l}} \prod_{j=1,j\neq l}^{L}\left(1+\frac{\gamma_l F_{lj}p_j}{p_l}\right)^{-1}\leq \bar{O}_l.
\end{align}
Taking the logarithm of \eqref{eq:outage1}, we have:
\begin{align}\label{eq:rif}
I_l(\p)=\frac{v_l + p_l\sum_{j=1,j\neq l}^{L}\log\left(1+\frac{\gamma_l F_{lj}p_j}{p_l}\right)}{\log(\frac{1}{1-\bar{O}_l})}\leq p_l,
\end{align}
where $I_l(\p)$ is proved to be a standard interference function in \cite{papandriopoulos2005optimal} and its concavity established later in \cite{tan2015optimal,zheng2016wireless}. Besides facilitating the algorithm design development of using $I_l(\p)$ in \eqref{eq:rif} to solve a broader class of utility maximization problems in \eqref{eq:prob_cvx} \cite{papandriopoulos2005optimal,boyd02,zheng2016wireless}, the deterministic form of $I_l(\p)$ in \eqref{eq:rif} gives the ground truth to evaluate the generalization error of the LLCP in \eqref{opt:cvxL} when fitting an unknown standard interference function. Specifically, we fit the standard interference function in \eqref{eq:rif} using the maximum monomials through \eqref{opt:cvxL}, which serves as a reference point to validate our approach.

\begin{example}\label{emp:emp2}
Consider a wireless network with two users whose transmit powers are subject to Rayleigh fading with SINR thresholds being $\gamma_1 =\gamma_2= 10$ and the noise variance $v_1 =v_2=0.1$ pW. The channel gain matrix is $\bm{G} = [1.19,2.13;1.69,1.64]^{\top}$. Let the outage constraints be $\bar{O}_1 = \bar{O}_2=0.1$. We use the LLCP model in\eqref{eq:mv} to approximate \eqref{eq:rif}. Setting $K=3$ in \eqref{eq:mv}, we obtain $\bm{c}_1=[2.1,1.6,6.3]$, $\bm{c}_2=[1.1,1.5,2.5]$, $\bm{c}_3=[0.1,1.2,1.3]$, $\bm{\theta}_1=[0.23,0.12;0.50,0.49]^{\top}$, $\bm{\theta}_2=[0.63,0.16;0.27,0.46]^{\top}$, and $\bm{\theta}_3=[0.47,0.25;0.57,0.26]^{\top}$. Fig.~\ref{fig:LLCP-fitting}(a) and Fig.~\ref{fig:LLCP-fitting}(b) show the effectiveness of LLCP in \eqref{eq:mv} to model outage probabilities under Rayleigh fading for User $1$ and User $2$. Blue scatter points show sample data; surfaces depict the approximated standard interference functions. LLCP-based monomials show a strong fit to the observed outage probabilities.
\end{example}

\subsubsection{Modeling Interference Function for Outage Constraints under Ricean Fading}

If the random variables characterizing the fading channel do not follow the independent Rayleigh distribution, but instead adhere to the Ricean or Nakagami distribution \cite{spacex10,spacex11}, obtaining expressions for outage probabilities in a deterministic manner becomes challenging, and thus using Lemma \ref{lemma2} directly to obtain the fixed points of the implicit standard interference functions for the utility maximization problem in \eqref{eq:prob_cvx} becomes impractical. Under Ricean fading, the received signal envelope has a Rayleigh distribution, and the received signal power has a non-central chi-square distribution \cite{stuber1996principles,spacex10,spacex11}:
\begin{align*}
   &\mathscr{F}(h_{lj}|K,\Omega)\\
=&\frac{K+1}{\Omega}\exp{(-K-\frac{(K+1)h_{lj}}{\Omega})}\mathscr{I}_0(2\sqrt{\frac{K(K+1)h_{lj}}{\Omega}})\;, 
\end{align*}
where $\mathscr{I}_0(\cdot)$ is the zeroth-order modified Bessel function of the first kind. As a special case with $K = 0$, Rician fading reduces to Rayleigh fading, which is known to yield a concave standard interference function \cite{tan2015optimal}. We approximate ${I}_l(\p)$ with monomial-based standard interference functions via the LLCP method, enabling the iterative algorithm in Section~\ref{sec:sec3} to compute approximate solutions.

\begin{figure}[htbp]
\centering

\subfigure[]{
\begin{minipage}[t]{0.5\linewidth}
\centering
\includegraphics[width=1.6in]{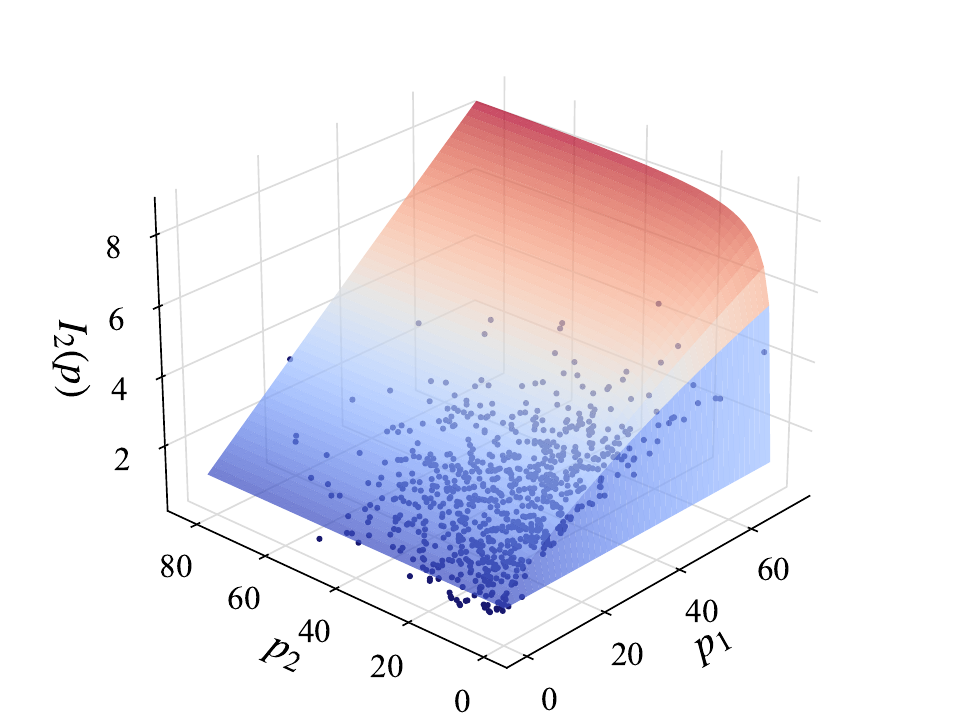}
\end{minipage}%
}%
\subfigure[]{
\begin{minipage}[t]{0.5\linewidth}
\centering
\includegraphics[width=1.6in]{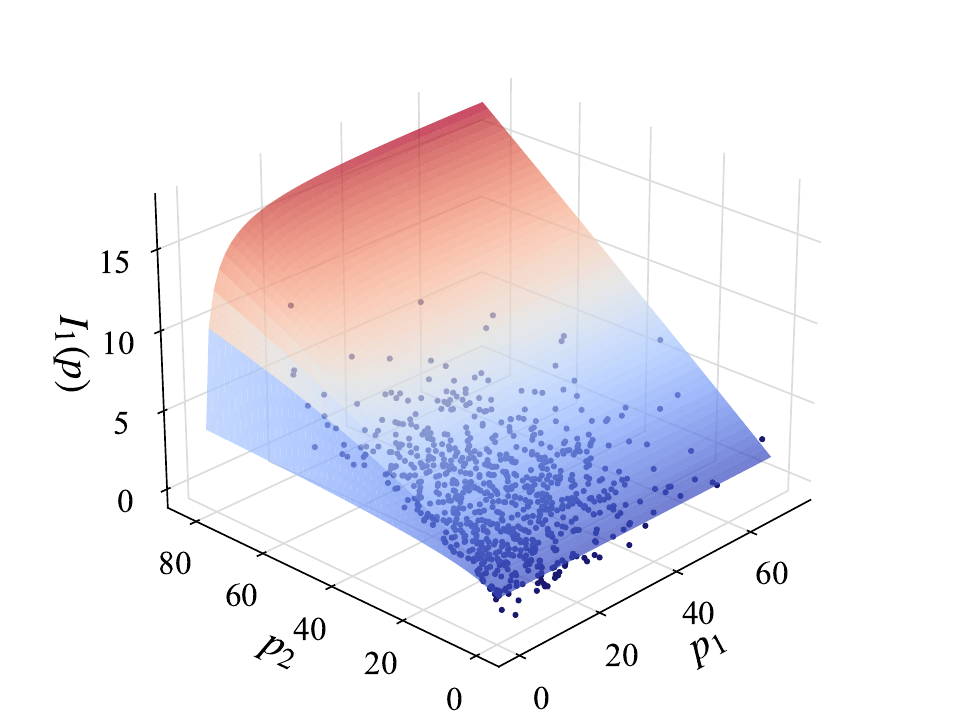}
\end{minipage}%
}%

\subfigure[]{
\begin{minipage}[t]{0.5\linewidth}
\centering
\includegraphics[width=1.6in]{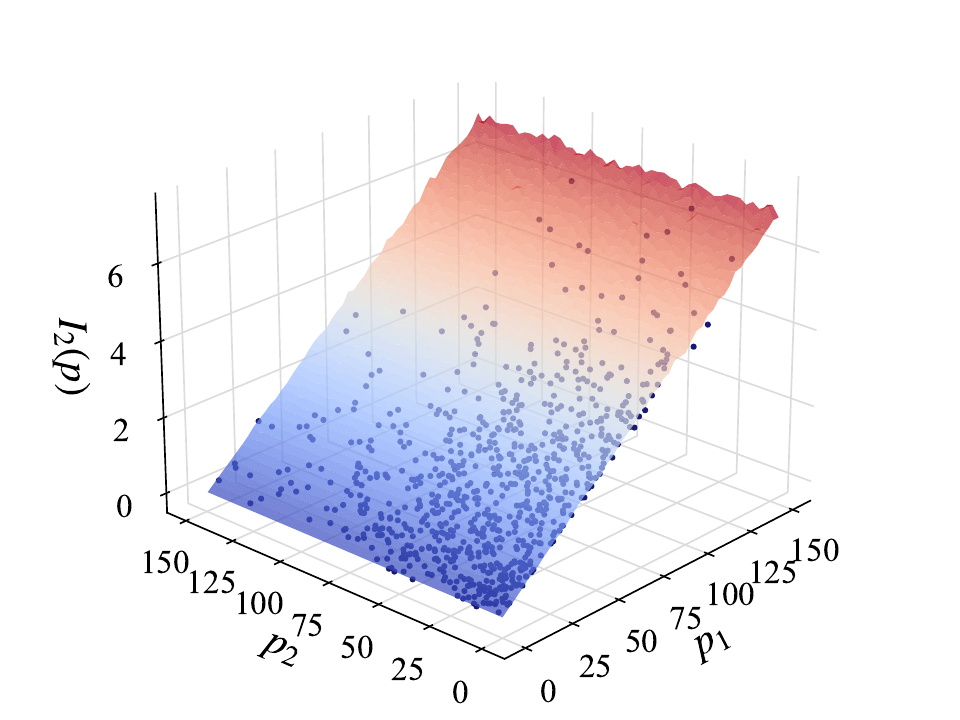}
\end{minipage}%
}%
\subfigure[]{
\begin{minipage}[t]{0.5\linewidth}
\centering
\includegraphics[width=1.6in]{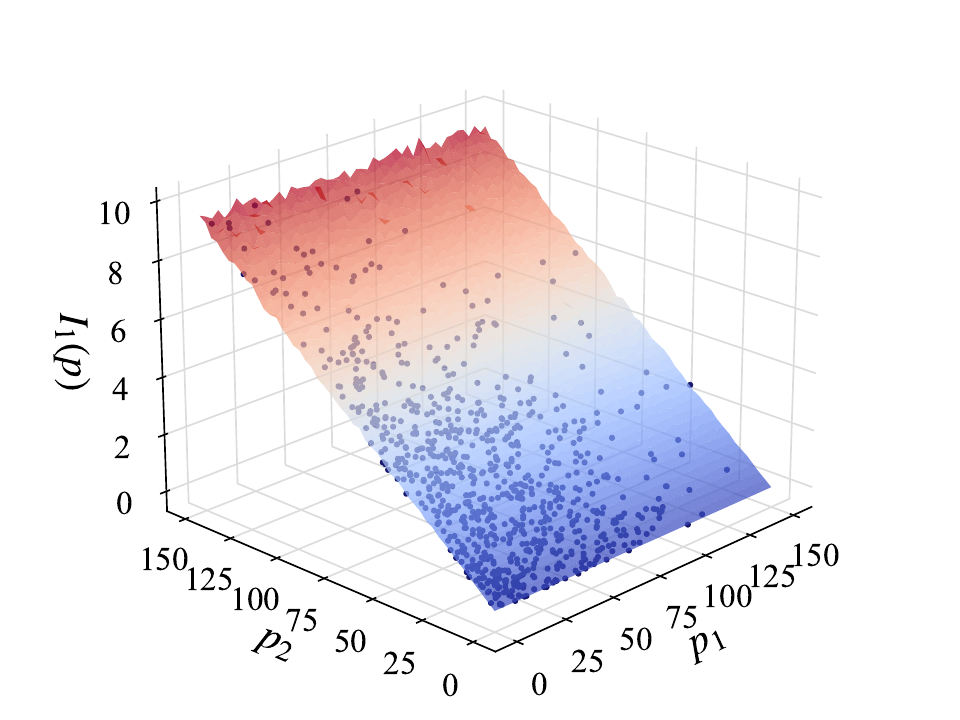}
\end{minipage}%
}%

\centering
\captionsetup{font={footnotesize}, justification=raggedright}
\caption{Illustration of using LLCP to approximate the outage probability constraints with monomials under Rayleigh fading ((a) and (b)) and Ricean fading ((c) and (d)). 
}
\label{fig:LLCP-fitting}
\end{figure}

\begin{example}
Let us consider another wireless network where the transmit powers are subject to Ricean fading. Given the SINR threshold $\gamma_i$ for each user, the outage constraint $\bar{O}_i$, the noise variance $v_i$, and the channel gain matrix as detailed in Example \ref{emp:emp2}, the sampled data generated by ${I}_l(\p)$ in \eqref{eq:rif} is depicted in Fig. \ref{fig:LLCP-fitting}. Using LLCP-based monomials to model the standard interference function ${I}_l(\p)$, the parameters in the approximation $\I(\p)$ in \eqref{eq:mv} are as follows: 
$\bm{c}_1=[1.8,0.9,1.6]$, $\bm{c}_2=[1.8,0.8,1.4]$, and $\bm{c}_3=[1.9,0.8,1.5]$, $\bm{\theta}_1=[0.13,0.43;0.65,0.29]^{\top}$, $\bm{\theta}_2=[0.12,0.42;0.67,0.30]^{\top}$, and $\bm{\theta}_3=[0.13,0.40;0.70,0.28]^{\top}$.
Fig.~\ref{fig:LLCP-fitting}(c) and Fig.~\ref{fig:LLCP-fitting}(d) show the effectiveness of LLCP in \eqref{eq:mv} to model the $\I(\p)$ in \eqref{eq:con_out} under Ricean fading for users $1$ and $2$. The blue scatter points represent the sample data, while the surfaces illustrate the approximate standard interference functions, also exhibiting an excellent fit.
\end{example}

\vspace{-2mm}

\subsection{Differentiable Programming by Algorithm Unrolling}
Algorithm unrolling is a deep learning technique that establishes a link between iterative algorithms (e.g., sparse coding) and neural network architectures \cite{gregor2010learning,monga2021algorithm} . In essence, algorithm unrolling maps each iteration to a network layer, enabling a trainable deep neural network that embeds optimization logic within a differentiable programming framework, enhancing both performance and interpretability \cite{edpbook}.

As an illustrative example, consider the stochastic outage probability problem under fading channel conditions:
\begin{align}\label{eq:prob}
\mbox{minimize}\;\;\;\; & -u(\p) \nonumber\\
\mbox{subject to}\;\;\;\; & \textup{Prob}(\sinr_l(\p)<\gamma_l)\leq \bar{O}_l \; \forall \; l.
\end{align}

Applying the LLCP model in \eqref{opt:cvxL} from Section \ref{sec:sec4} to the outage probability constraints, we obtain an approximation of problem \eqref{eq:prob}:
\begin{align}\label{eq:prob_unroll}
\mbox{minimize}\;\;\;\; & -u({\p}) \nonumber\\
\mbox{subject to}\;\;\;\; & \max_{k=1,\cdots,K}\left\{c_{kl}\prod_{j=1}^{L}p_j^{\theta_{klj}}\right\}\leq {p}_l \; \forall \; l.
\end{align}

According to Theorem \ref{thm:thm1}, the optimal solution to the problem \eqref{eq:prob_unroll} can be obtained by the following iteration:
\begin{align}\label{eq:eq3}
    {p}_l(t+1) =  \max_{k=1,\cdots,K}\left\{c_{kl}\prod_{j=1}^{L}p_j^{\theta_{klj}}(t)\right\},
\end{align}

\begin{align}\label{eq:eq4}
   & x_l(t+1) = -\frac{\partial u(\p(t))}{\partial p_l}+c_{kl}\prod_{i=1}^L{p}_i^{\theta_{kli}}(t)\sum_{j=1}^L \theta_{klj}p_j^{-1}(t)x_j(t),
\end{align}
where $k=\textup{argmax}_{\hat{k}} c_{\hat{k}l}\prod_{j=1}^{L}p_j^{\theta_{\hat{k}lj}}(t)$ for $\hat{k}=1,\dots,K$,\\
and
\begin{align}\label{eq:eq5}
   {\lambda}_l(t+1) = p_l(t+1) x_l(t+1).
\end{align}
 
We now apply algorithm unrolling to construct a fixed-depth nonlinear feedforward model, trained to approximate the optimal solution of \eqref{eq:prob}. The key idea is to utilize feedforward networks whose structure corresponds to a finite number of iteration steps in \eqref{eq:eq3}, \eqref{eq:eq4}, and \eqref{eq:eq5}. Denote the model depth by $T$. We let $\tilde{\p} = \log \p$ and $\tilde{\bm{c}} = \log \bm{c}$. The architecture can be interpreted as a time-unfolded recurrent neural network over steps $t \leq T$, as follows:
\begin{align}
&\label{eq:unrolling1}\bm{s}_k(t\!+\!1)=\bm{f}\left(\Hat{\mathbf{w}}(t\!+\!1)(\bm{\theta}_k\tilde{\p}(t)+\tilde{\bm{c}}_k)\right),\quad\\
    &\tilde{\p}(t\!+\!1)=\max_{k=1,\cdots,K}\left\{\bm{s}_k(t\!+\!1)\right\},\quad\\
    &\x(t+1)=\bm{f}\left(\breve{\mathbf{w}}(t)\nabla(\I(\p(t\!+\!1)))\x(t)+\nabla u(\p(t\!+\!1))\right),\\
    &\label{eq:unrolling4}\bm{\lambda}(t+1) = \p(t+1)\circ \x(t+1),
\end{align}
where $\bm{f}(\cdot)$ is a nonlinear activation function (e.g., ReLU) and $$
\nabla(\I(\p(t)))_{lj}=c_{kl}\prod_{i=1}^L{p}_i^{\theta_{kli}}(t)\theta_{klj}p_j^{-1}(t),
$$ 
where $k=\textup{argmax}_{\hat{k}} c_{\hat{k}l}\prod_{j=1}^{L}p_j^{\theta_{\hat{k}lj}}(t)$ for $\hat{k}=1,\dots,K$. Denote the architecture of our unrolled model by $[\p(T);\bm{\lambda}(T)]=\bm{\mathcal{F}}(\bm{\theta},\bm{c},\mathbf{w})$, where $\mathbf{w} = \{\Hat{\mathbf{w}}(t),\breve{\mathbf{w}}(t)\}$ for $t=1,\dots,T$. Suppose we are given a training set $\Bar{\bm{\mathcal{D}}}$ with $|\Bar{\bm{\mathcal{D}}}|$ training samples. For the $m$th sample $\{(\bm{\theta}_m,\bm{c}_m),(\p_m^*,\bm{\lambda}_m^*)\}\in \Bar{\bm{\mathcal{D}}}$,  $(\bm{\theta}_m,\bm{c}_m)$ are the inputs and $(\p^*,\bm{\lambda}^*)$ are the ground truth of outputs. The loss function is defined as the squared error between the predicted solution and the optimal solution to the problem \eqref{eq:prob_unroll}:
\begin{align}\label{eq:sif_loss}
    E(\mathbf{w}) = \frac{1}{|\Bar{\bm{\mathcal{D}}}|}\!\!\!\!\sum_{\{(\bm{\theta}_m,\bm{c}_m),(\p_m^*,\bm{\lambda}_m^*)\}\in \Bar{\bm{\mathcal{D}}}}\!\!\!\!\!\!\!\!\|[\p_m^*;\bm{\lambda}_m^*]-\bm{f}(\bm{\theta}_m,\bm{c}_m,\mathbf{w})\|_2.
\end{align}
We train the parameters $\mathbf{w}$ by minimizing the loss $E(\mathbf{w})$, with gradients obtained via automatic differentiation (e.g., {\tt PyTorch autograd}). Algorithm \ref{alg:deepsif} presents DIFFRACT for computing approximate primal and dual solutions to \eqref{eq:prob_unroll}.

\begin{algorithm}
\SetAlgoLined
\textbf{Input}: Training dataset $\Bar{\bm{\mathcal{D}}}$ with input-output pairs $\{\mathcal{D}_m,(\p_m^*,\bm{\lambda}_m^*)\}\in\Bar{\bm{\mathcal{D}}}$, where $\mathcal{D}_m$ is a dataset with input-output pairs $(\p_m{(i)},\y_m{(i)})\in \mathcal{D}_m.$\\
\textbf{Initialize:} $\p(0)\!=\!\bm{0},\!\x(0)\!=\!\bm{0}$, learning parameters  $\mathbf{w}(0)\! =\! (\mathbf{w}_1(0),$ $\mathbf{w}_2(0),$ $\bm{b}_1(0),$ $\bm{b}_2(0))$ and learning rate $\eta_w$.\\
\For{$m = 1,\dots,|\Bar{\bm{\mathcal{D}}}|$}{
\!\!Obtain $\bm{c}_m$ and $\bm{\theta}_m$ of the $|\Bar{\bm{\mathcal{D}}}|$ monomial vectors using Algorithm \ref{alg:alg1} with $\mathcal{D}_m$.\\
\!\!\For{$t = 1,\dots,T$}{
\begin{align*}
&\bm{s}_k(t\!+\!1)=\bm{f}\left(\Hat{\mathbf{w}}(t\!+\!1)(\bm{\theta}_k\tilde{\p}(t)+\tilde{\bm{c}}_k)\right),\quad\\
    &\tilde{\p}(t\!+\!1)=\max_{k=1,\cdots,K}\left\{\bm{s}_k(t\!+\!1)\right\},\quad\\
    &\x(t\!\!+\!\!1)\!=\!\bm{f}\left(\breve{\mathbf{w}}(t)\nabla(\I(\p(t\!\!+\!\!1)))\x(t)\!\!+\!\!\nabla u(\p(t\!+\!1))\right),\\
    &\bm{\lambda}(t+1) = \p(t+1)\circ \x(t+1).
\end{align*}
}
Use automatic differentiation (e.g., {\tt Pytorch autograd}) to compute the gradient $\nabla E(\mathbf{w}(t))$ and update $\mathbf{w}(t+1) $ $= \mathbf{w}(t)$ $-\eta_w\nabla E(\mathbf{w}(t))$.
}
\caption{DIFFRACT via Algorithm Unrolling}
\label{alg:deepsif}
\end{algorithm}

Implemented in PyTorch, Algorithm \ref{alg:deepsif} frames the primal-dual optimization loop as a differentiable computational graph. Each iteration is modeled as a neural layer using learned monomial mappings for both primal and dual updates. Leveraging PyTorch's autograd engine in \cite{agrawal2019differentiable}, this setup can enable end-to-end training with gradient-based optimization and GPU acceleration integrated with NVIDIA's SONNIA framework \cite{hoydis2023sionna,hoydis2024learning} by integrating Algorithm \ref{alg:deepsif} with $\nabla$-Prox \cite{deltaprox2023}, an ML compilation framework for generating memory-efficient differentiable solvers, thus enabling rapid prototyping and fully differentiable resource allocation under interference constraints in an AI-native wireless network stack.


\section{Performance Evaluation}
In this section, we provide numerical examples to demonstrate the effectiveness of the proposed DIFFRACT framework for solving the utility maximization problem in Section \ref{sec:sec3}. 


\subsection{Performance of LLCP}
Consider a wireless network with eight transmitter-receiver pairs under Rayleigh, Ricean, and Nakagami fading, focusing on minimizing total power while meeting outage probability constraints in \eqref{eq:prob}. For each scenario, we generate 1,000 samples of the implicit standard interference function derived from the outage probability constraints. We then apply the LLCP model to fit this sample data, allowing us to approximate the implicit standard interference functions using a maximum of monomials for each problem instance.
The LLCP model for fitting the implicit standard interference functions is set as \eqref{eq:mv}, where $c_{kl}$ and $\theta_{klj}$ for all $k$, $l$ and $j$ are learning parameters. We set the number of monomials as $K=3,5,8$ and $10$. The loss function is defined as \eqref{eq:loss}, where the penalty scalar $M=1\times 10^{3}$. We set the learning rate as $2\times10^{-3}$.  

 To assess the performance of the LLCP model in \eqref{opt:cvxL}, we use metrics: utility accuracy and power allocation accuracy. Utility accuracy measures the ratio between the predicted utility and the maximum achievable utility, indicating the model's ability to approximate optimal utility values. Power allocation accuracy measures the closeness of predicted to ground truth allocations, reflecting the model's effectiveness in solving the problem. We approximate the standard interference function using Lemma \ref{lemma5}. For Rayleigh fading, ground truth values come from the iterative algorithm in \cite{tan2015optimal}, while for Rician and Nakagami fading, we use exhaustive search for benchmarking. Tables~\ref{tab:uc} and \ref{tab:pc} compare the utility and power allocation accuracy using different numbers of monomials to fit implicit standard interference functions under various fading conditions. We observe in our numerical evaluations that the LLCP model approximates these functions well at $K=8$ and only marginal gains beyond that.

\subsection{Performance of Algorithm Unrolling}
We now evaluate the algorithm unrolling method for the approximate standard interference function using the maximum of monomials. We consider the utility maximization problem involving eight users, with the number of monomials fixed at $8$. The unrolled model for computing the maximum over monomials is given by \eqref{eq:unrolling1} to \eqref{eq:unrolling4}. The model takes as input the vectors $\bm{\theta}_k$ and $\mathbf{c}_k$, and is trained using the loss in \eqref{eq:sif_loss}. We generate $10^6$ samples, splitting them into $80\%$ for training and $20\%$ for testing, with 10-fold cross-validation for robustness.
The validation datasets are constructed using the approximate standard interference functions defined earlier. For problem instances subject to Rayleigh fading, ground truth values are obtained using the iterative algorithm in \cite{tan2015optimal}. For scenarios involving Ricean and Nakagami fading, we compute the ground truth using a brute force discretization search.

\begin{figure*}[htbp]
\centering
\subfigure[]{
\begin{minipage}[t]{0.3\linewidth}
\centering
\includegraphics[width=1.7in]{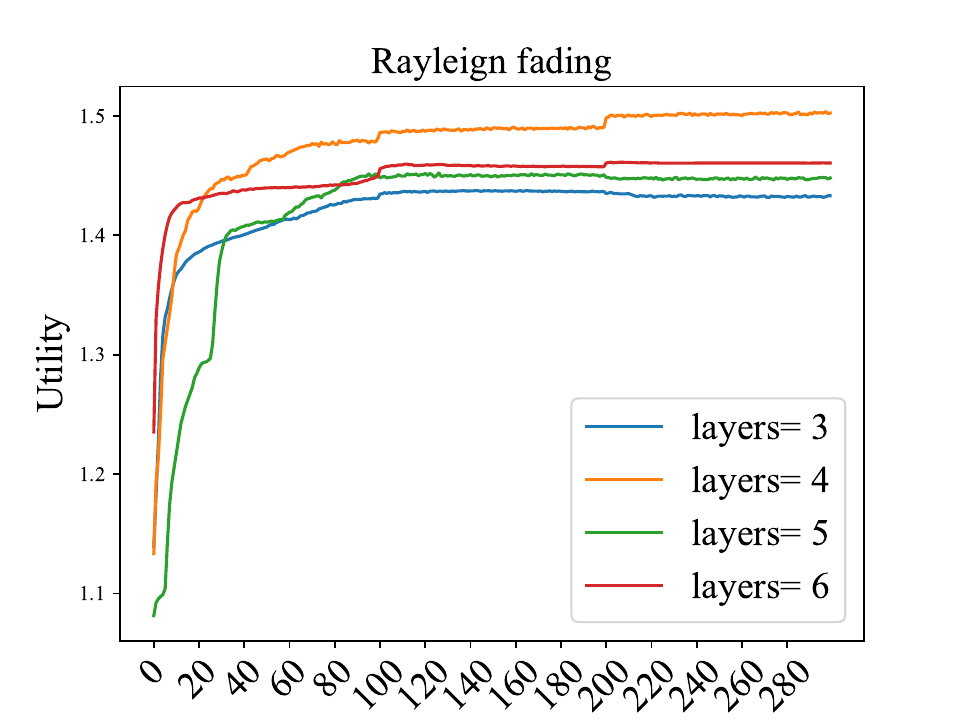}
\end{minipage}%
}%
\subfigure[]{
\begin{minipage}[t]{0.3\linewidth}
\centering
\includegraphics[width=1.7in]{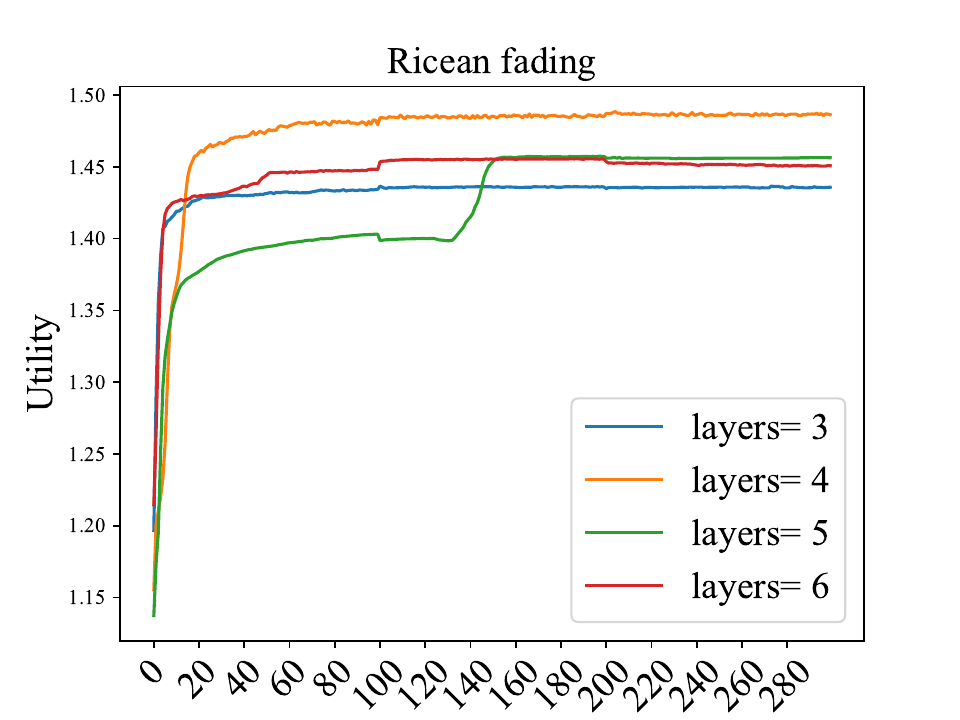}
\end{minipage}%
}%
\subfigure[]{
\begin{minipage}[t]{0.3\linewidth}
\centering
\includegraphics[width=1.7in]{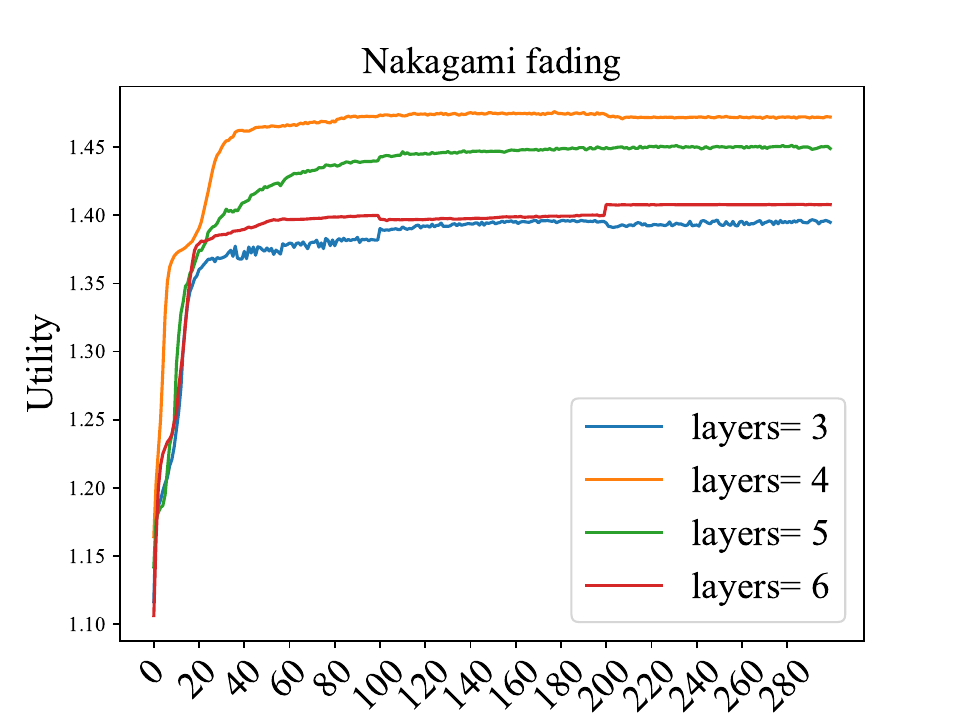}
\end{minipage}%
}
\centering
\captionsetup{font={footnotesize}, justification=raggedright}
\caption{An illustration of the performance of Algorithm 2 in unrolling the standard interference function to compute the optimal utility values of the utility maximization problem under various fading conditions: Rayleigh fading (a), Ricean fading (b), and Nakagami fading (c).}
\label{fig:res2}
\end{figure*}

We first examine how the number of layers in the unrolled model affects the approximation of standard interference functions for utility maximization under different fading environments. Fig. \ref{fig:res2} illustrates the maximum utility results obtained from the unrolled model across varying numbers of layers under Rayleigh fading, Ricean fading, and Nakagami fading, respectively. Interestingly, the model achieves the highest utilities with 4 layers. We also demonstrate the unrolled model's superior performance by comparing it against established baselines: the Adaptive Outage-based Power Control (AOPC) algorithm in \cite{tan2015optimal} for Rayleigh fading, the large deviations principle (LDP) algorithm from \cite{rao2024outage} for Ricean fading, and the convex relaxation (CR) algorithm from \cite{fischione2011utility} for Nakagami fading. TABLE \ref{tab:comp} summarizes the results of maximal utility under different fading environments. We observe that while our unrolled model yields slightly lower values than the ground truth, it outperforms other approximation methods. Table IV further shows that it achieves the highest efficiency compared to all baselines. TABLE \ref{tab:comp} presents the results of maximal utility across various fading environments. Although the output from our unrolled model are slightly below the ground truth values, they surpass the performance of other approximate methods. Furthermore, Table IV compares the efficiency of our proposed methods with several baselines, highlighting that the unrolled model achieves the highest efficiency.

\begin{table}[]
\small
\centering
\begin{tabular}{|c|c|c|c|}
\hline
 & Rayleigh Fading& Ricean fading & Nakagami fading \\ \hline
$K=3$       & 0.9291          & 0.9248        & 0.9235          \\ \hline
$K=5$       & 0.9391          & 0.93497        & 0.9400          \\ \hline
$K=8$       & 0.9834          & 0.9876        & 0.9793          \\ \hline
$K=10$      & 0.9856          & 0.9909        & 0.9834          \\ \hline
\end{tabular}
\centering
\captionsetup{font={footnotesize}, justification=raggedright}
\caption{Comparison of utility accuracy across various monomials used to fit implicit standard interference functions for wireless systems in diverse fading environments.}
\label{tab:uc}
\end{table}

\begin{table}[]
\small
\centering
\begin{tabular}{|c|c|c|c|}
\hline
 & Rayleigh Fading& Ricean fading & Nakagami fading \\ \hline
$K=3$       & 0.9177          & 0.9131        & 0.9195          \\ \hline
$K=5$       & 0.9259          & 0.9302        & 0.9228          \\ \hline
$K=8$       & 0.9697          & 0.9617       & 0.9645          \\ \hline
$K=10$      & 0.9760          & 0.9633        & 0.9687          \\ \hline
\end{tabular}
\centering
\captionsetup{font={footnotesize}, justification=raggedright}
\caption{Comparison of power accuracy across various monomials used to fit implicit standard interference functions for wireless systems in diverse fading environments.}
\label{tab:pc}
\end{table}

\begin{table}[]
\footnotesize
\centering
\begin{tabular}{|c|c|c|c|}
\hline
Environment                                     & Rayleigh fading & Ricean fading & Nakagami fading \\ \hline
Algorithm 1                                     & 0.9902          & 0.98851       & 0.9899          \\ \hline
Algorithm 2                                     & 0.9704          & 0.9622        & 0.9664          \\ \hline
AOPC \cite{tan2015optimal}      & 1.0000          & --            & --              \\ \hline
LDP \cite{rao2024outage}       & --              & 0.9525        & --              \\ \hline
CR \cite{fischione2011utility} & --              & --            & 0.9434          \\ \hline
\end{tabular}
\centering
\captionsetup{font={footnotesize}, justification=raggedright}
\caption{Comparison of the utility accuracy performance for approximate standard interference functions using LLCP (Algorithm 1) and Algorithm Unrolling (Algorithm 2), benchmarked against different baselines based on existing state-of-the-art algorithms for various wireless fading conditions.}
\label{tab:comp}
\end{table}

\begin{table}[]
\footnotesize
\centering
\begin{tabular}{|c|c|c|c|}
\hline
Environment                                     & Rayleigh fading & Ricean fading & Nakagami fading \\ \hline
Algorithm 1                                     & 10.9086          & 11.2013      & 11.3778          \\ \hline
Algorithm 2                                     & 1.5690          & 1.6889        & 1.2701          \\ \hline
AOPC \cite{tan2015optimal}      & 9.5252          & --            & --              \\ \hline
LDP \cite{rao2024outage}       & --              & 8.6566        & --              \\ \hline
CR \cite{fischione2011utility} & --              & --            & 9.6425          \\ \hline
\end{tabular}
\centering
\captionsetup{font={footnotesize}, justification=raggedright}
\caption{Comparison of time efficiency (seconds) for approximate standard interference functions using LLCP (Algorithm 1) and Algorithm Unrolling (Algorithm 2), benchmarked against different baselines under various fading channel conditions.}
\label{tab:comp}
\end{table}

\section{Conclusion}
This paper presents DIFFRACT, a general differentiable programming framework for utility maximization in wireless networks, particularly effective in managing resources under unpredictable interference, as commonly encountered in satellite-terrestrial systems. Our work reveals new duality results by establishing the differentiability and monotone operator characterization of standard interference functions. This provides a nonlinear Perron-Frobenius operator perspective that enables smooth integration into differentiable programming and modern deep learning frameworks with efficient automatic differentiation. DIFFRACT models complex interference patterns--such as outage in atmospheric fading, which is critical for space-to-ground communication--within an \textit{end-to-end trainable, AI-native stack} by unrolling a utility maximization algorithm with implicit interference constraints into differentiable primal-dual updates. Experiments show DIFFRACT offers theoretically sound performance and scalability for AI-native resource control in wireless network optimization. Future work will leverage the monotone operator characteristics in Theorems \ref{lemma4} to \ref{npftheorem} to design deep implicit layers in the DIFFRACT framework, enabling differentiation through utility maximization via fixed-point strategies \cite{combettes2021fixed,monotoneoperatornetworks}. This approach will integrate first-order neural acceleration to improve convergence and enable scalable, distributed learning in AI-native wireless networks. 

\section{Acknowledgement}
The research was supported in part by NTU startup and the Singapore Ministry of Education Academic Research Fund (MOE-T2EP20224-0009).

\bibliographystyle{IEEEtran}
\bibliography{_Reference}

\end{document}